\documentclass[iop,apj]{emulateapj}
\slugcomment{}

\shorttitle{DISK FORMATION AND GRAVITATIONAL EVOLUTION}
\shortauthors{TAKAHASHI, INUTSUKA, \& MACHIDA}

\usepackage{natbib, aas_macros}
\begin{document}


\title{A Semi-Analytical Description for the Formation and
 Gravitational Evolution of Protoplanetary Disks}


\author{Sanemichi Z. Takahashi\altaffilmark{1,2}, Shu-ichiro
Inutsuka\altaffilmark{1}, and Masahiro N. Machida\altaffilmark{3}}
\altaffiltext{1}{Department of Physics, Nagoya University, Furo-cho, 
Chikusa-ku, Nagoya, Aichi, 464-8602, Japan;
takahashi.sanemichi@a.mbox.nagoya-u.ac.jp, inutsuka@nagoya-u.jp
}
\altaffiltext{2}{Department of Physics, Kyoto University, Oiwake-cho,
Kitashirakawa, Sakyo-ku, Kyoto 606-8502, Japan;
sanemichi@tap.scphys.kyoto-u.ac.jp}
\altaffiltext{3}{Department of Earth and Planetary Science, Kyushu
University, Higashi-ku, Fukuoka 812-8581, Japan; machida.masahiro.018@m.kyushu-u.ac.jp}



\def\bm#1{\mbox{\boldmath $#1$}}

\begin{abstract}
We investigate the formation process of self-gravitating protoplanetary
disks in unmagnetized molecular clouds. 
The angular momentum is redistributed by the action
of gravitational torques in the massive disk during its early formation. 
We develop a simplified one-dimensional accretion disk model
that takes into account the infall of gas from the envelope onto the
disk and the transfer of angular momentum in the disk with
an effective viscosity. 
First we evaluate the gas accretion rate from the cloud core onto
 the disk by approximately estimating the effects of gas pressure and
 gravity acting on the cloud core. 
 We formulate the effective viscosity
 as a function of the Toomre $Q$ parameter
that measures the local gravitational stability of the rotating thin
 disk. 
We use a function for viscosity that changes sensitively with $Q$ 
when the disk is  gravitationally unstable. 
We find a strong self-regulation mechanism in the disk evolution. 
During the formation stage of protoplanetary disks, the evolution of 
the surface density does not depend on the other 
details of the modeling of effective
 viscosity, such as the prefactor of the viscosity coefficient.
Next, to verify our model,
we compare the time evolution of the disk calculated with our
 formulation with that of three-dimensional hydrodynamical
 simulations. The structures of the resultant disks from the one-dimensional
 accretion disk model agree well with those of the three-dimensional
 simulations. 
Our model is a useful tool for the further modeling of chemistry, radiative
 transfer, and planet formation in protoplanetary disks.
\end{abstract}



\section{Introduction}
Since planets are expected to form in protoplanetary disks, 
planet-formation scenarios should depend on the
structure of protoplanetary disks formed through realistic star
formation processes. 
Recent observations have revealed the planets in wide orbits of more than 30 AU
\cite[]{2008Sci...322.1345K,2008Sci...322.1348M,2009ApJ...707L.123T}.
The most convincing scenario to form such planets is the fragmentation of
disks due to gravitational instability
\cite[]{2007MNRAS.382L..30S,2009ApJ...707...79D,2010ApJ...714L.133V,2010ApJ...724.1006M}.
To investigate the gravitational instability of protoplanetary disks, 
we need the temperature and density structure of disks.
These disk properties can be obtained by revealing the formation
process of protoplanetary disks.  
\cite{1981Icar...48..353C} have done pioneering theoretical work on the
formation and evolution of protoplanetary disks.
They calculated the gravitational collapse of a cloud
core during the formation of a protoplanetary disk in which they
treated the viscosity coefficient of the disk as a parameter.
Their modeling was done before more realistic three-dimensional
simulations became available.

Recent three-dimensional non-ideal MHD numerical simulations suggest that protoplanetary disks
are gravitationally unstable in their early formation stages because
the masses of the disks remain very large
\cite[]{2010ApJ...718L..58I,2010ApJ...724.1006M,2011PASJ...63..555M}.
Massive disks are also formed with non-MHD thin-disk
calculations \cite[e.g.,][]{2006ApJ...650..956V,2011ApJ...729..146V}.
In such disks, spiral arms are formed in the case that 
$1\lesssim Q\lesssim 2$, where  $Q\equiv \kappa c_{\rm{s}}/\pi G\Sigma $
is Toomre's parameter, $\kappa$ is the epicyclic frequency, $c_{\rm{s}}$ is
the sound speed, $G$ is the gravitational constant, and $\Sigma$ is the 
surface density of the disk.
The angular momentum in disks is redistributed by the action of
gravitational torques due to spiral arms.
There are some effective viscosity models to mimic the angular
momentum transfer due to gravitational torque.
An $\alpha$-prescription \cite[]{1973A&A....24..337S} is used for the
effective viscosity models, $\nu_{\rm{eff}}=\alpha c_s^2/\Omega$, where
$\Omega$ is the angular frequency of a disk, $\nu_{\rm{eff}}$
is the effective viscosity, and $\alpha$ is a non-dimensional parameter.
\cite{1987MNRAS.225..607L} and \cite{2008ApJ...681..375K} investigated
the functional forms of $\alpha$ during disk evolution.
\cite{1994ApJ...421..640N,1995ApJ...445..330N} and \cite{2010ApJ...713.1143Z}
performed one-dimensional numerical simulations on the formation of
protoplanetary disks from the collapse of a cloud cores using an
effective viscosity.
Various effective viscosity models were also used in numerical simulations of
 massive star formation  \cite[]{2002ApJ...569..846Y,2012ApJ...760L..37H}.
These are axisymmetric two-dimensional simulations using
an effective viscosity in place of time-consuming three-dimensional simulations.
However, an effective viscosity model that can mimic angular momentum
transfer due to gravitational torque is still unknown.
\cite{2010NewA...15...24V} calculated the formation of protoplanetary
disks using the 
effective viscosity models suggested in \cite{1987MNRAS.225..607L} and
\cite{2008ApJ...681..375K}, and compared the resultant disks and
protostars with the results of two-dimensional numerical simulations. 
He concluded that both models cannot mimic the
gravitational torque. 
Thus, further investigation is needed to construct a realistic
 effective viscosity model, especially when the disk is sufficiently massive to be
 gravitationally unstable.

In this work, we perform both one-dimensional numerical
calculations and three-dimensional numerical simulations.
We model the accretion disks using effective viscosity models in the
one-dimensional calculation and calculate the formation and evolution of
protoplanetary disks in the simulations.
Using our results we construct effective viscosity models
that can mimic the gravitational torques in the three-dimensional simulations. 
We also investigate the properties of the resultant disks using our 
one-dimensional calculations with a wide range of parameters.

We describe our one-dimensional model that incorporates
 effective viscosity and gas
infall from the cloud core in Section \ref{sec:eqs}. 
In Section \ref{sec:res}, we investigate the resultant disks of our
calculations. 
In Section \ref{sec:3D}, we describe our three-dimensional
numerical simulation and show the comparison of the effective viscosity
models with the three-dimensional simulations.
In Section \ref{sec:disc}, we discuss the properties of the resultant
disks.
We summarize our main conclusions in Section \ref{sec:conc}.

\section{Basic Equations and Settings for One-dimensional Accretion Disk
 Model}
\label{sec:eqs}
\subsection{Initial Conditions and Accretion onto Disk}
Recent observations suggest that the density structures of some prestellar
molecular cloud cores
can be approximated by Bonnor-Ebert spheres \citep[e.g.,][]{2001Natur.409..159A}.
As an initial state, we adopt a Bonnor-Ebert density profile \cite[]{1955ZA.....37..217E,1956MNRAS.116..351B} with central
density $3\times 10^5\ \rm{cm^{-3}}$, radius $17400\rm{AU}$ 
and temperature $10 \rm{K}$.
We increase the density by a factor of $f=$1.4 to promote gravitational
collapse.
The resultant cloud mass is $2.5M_{\rm \odot}$.
The cloud core is initially in uniform rotation with angular frequency $\Omega_0 = 4.8\times10^{-14}\ \rm{s^{-1}}$.
 To investigate the cloud evolution, the ratios of
thermal and rotational energy to the gravitational energy of the initial
cloud are useful and have been used in many previous studies
\citep[]{1999ApJ...523L.155T, 2003ApJ...595..913M, 2010ApJ...724.1006M}.  
The parameters $f$ and $\Omega_0$ are related to them as 
\begin{equation}
 \frac{U}{|W|} = 0.7f^{-1}, 
\end{equation}
\begin{equation}
 \frac{T}{|W|} = 1.2 \times 10^{-2} f^{-1}
  \left(\frac{\Omega_0}{4.8 \times 10^{-14}\ {\rm s^{-1}}}\right)^2,
\end{equation}
where $W$ is gravitational energy, $U$ is thermal energy, and $T$ is kinetic
energy in the initial core. 
 Note that these parameters are not constants in the time evolution.

\begin{figure}
\epsscale{1}\plotone{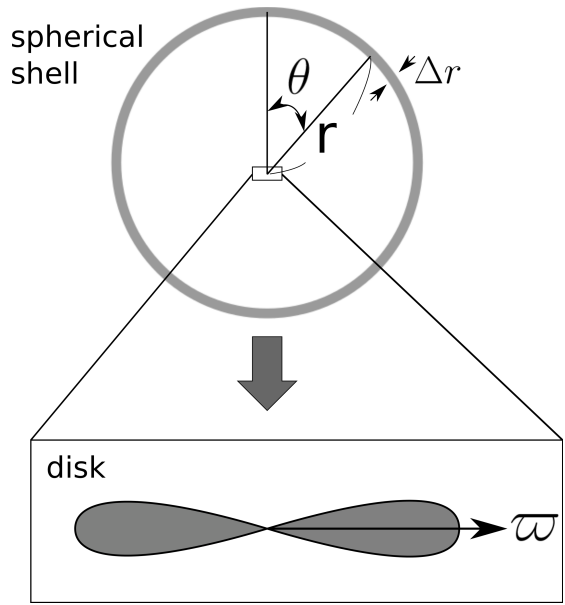}
  \caption{Schematic picture for spherical shell and disk formed at the
 center, where $r$ is the radius from the center of the cloud core,
 $\theta$ is the angle between the initial position of the gas
 and the rotation axis, $\Delta r$ is the thickness of the spherical
 shell, and $\varpi$ is the radial coordinate of the disk.
}
  \label{fig:cloud-disk}
 \end{figure}

Previous studies adopted the density of cloud cores as $\rho
\propto r^{-2}$ \citep[e.g.,][]{1981Icar...48..353C}.
In this case, the mass accretion rate onto the disk is constant
\cite[]{1977ApJ...214..488S}. In our work, a Bonnor-Ebert density profile
is adopted as an initial condition 
so that we have to model the mass accretion rate from
Bonnor-Ebert spheres, which depend on time.
We divide the cloud core into spherical shells of thickness $\Delta r$
and consider the collapsing motion of each shell (Figure \ref{fig:cloud-disk}).
From the equation of motion, the velocity of the shells, $u$, obeys the
following equation of motion: 
\begin{eqnarray}
 \frac{Du}{Dt}&=&-\frac{c_{\rm{s}}^2}{\rho}\frac{\partial \rho}{\partial r}
-\frac{GM_{\rm in}}{r^2}\nonumber\\
 &=&\frac{c_{\rm{s}}^2}{r}{F(r)}-\frac{GM_{\rm in}}{r^2},
\end{eqnarray}
where $r$ is the radius from the center of the cloud core, $M_{\rm in}$
is the total mass
within the shell, and $c_{\rm{s}}$ is the sound speed of the shell.
The total mass within the shell is constant during the collapse as
$M_{\rm in}=M_{r_{\rm{ini}}}$, where $r_{\rm{ini}}$ is the initial radius of
the shell and $M_{r_{\rm{ini}}}$ is the total mass initially contained
within the radius $r_{\rm ini}$.
Since the envelope is isothermal with $T=10$K, the sound speed is 
constant in the envelope.
The function $F(r)=(r/\rho)\partial \rho / \partial r$ depends on the density
profile and its value is expected to be of order unity.
Since the collapsing shells spend most of their time at outer radii,  we
approximate $F(r)$ at the initial radius as $F(r)=F(r_{\rm{ini}})=\rm{const}$.
The initial pressure is equal to the initial gravity before the mass is
enhanced by factor $f$,
\begin{eqnarray}
F(r_{\rm{ini}})& =& \frac{GM_{r_{\rm{ini}}}}{fc_{\rm{s}}^2r_{\rm{ini}}}.
\end{eqnarray}
From this approximation, the time in which the gas accretes from the cloud
onto the disk is expressed as 
\begin{equation}
 t_{\rm{infall}}= \sqrt{\frac{r_{\rm{ini}}}{2GM_{r_{\rm ini}}}}\int^1_0{\frac{dR}{\sqrt{\frac{1}{f}\ln R +
  \frac{1}{R}-1}}}.\label{eq:t_infall}
\end{equation}
Equation (\ref{eq:t_infall}) indicates that  $r_{\rm{ini}}$ is a function of
$t_{\rm{infall}}$ as  $r_{\rm ini}=r_{\rm{ini}}(t_{\rm{infall}})$.
Therefore, the initial radius of the shell that accretes onto the disk at
time $t$ is given by  $r_{\rm{ini}}(t)
$.
The thickness of the shell accreting onto the disk per unit time is
given by $dr_{\rm{ini}}/dt$.

\begin{figure}
\epsscale{1}\plotone{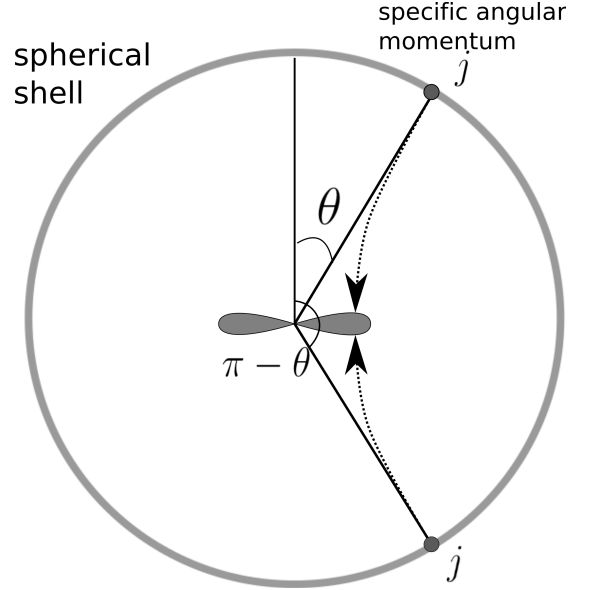}
  \caption{Schematic picture of gas accretion from the cloud core.
The specific angular momentum of gas with an initial angle
 $\theta=\theta_0 $ is the same as that with $\theta=\pi-\theta_0$.
 Thus these gases accrete onto the same
 region of the disk as defined by the centrifugal balance.
Note that the actual size of the disk is much smaller than shown in this
 schematic picture.
}
  \label{fig:cloud-disk2}
 \end{figure}

We assume that a spherical shell of the cloud core accretes onto the disk
almost simultaneously, and that angular momentum is conserved throughout this process
since axial symmetry is almost preserved. 
This assumption is justified  in three-dimensional
simulations without large initial non-axisymmetric perturbations 
\cite[e.g.,][]{2010ApJ...724.1006M}.
We assume that gas accretes onto the disk region where central gravitational force
is balanced by centrifugal force. In other words, the accretion radius
is determined by the balance between gravitational force and centrifugal force.
The mass accretion rate onto the disk from the cloud core per unit
radius is given by
$\partial\dot{M}_{\varpi\rm{,infall}}/\partial{\varpi}$,
where $\varpi$ is the radial coordinate of the disk, and
$\dot{M}_{\varpi\rm{,infall}}(\varpi, t)$ is the total mass of gas that accretes
onto the disk within radius $\varpi$ per unit
time. $\dot{M}_{\varpi\rm{,infall}}(\varpi, t)$ is equal to the total
mass of the accreting gas per unit time whose specific angular momentum is
smaller than $j(\varpi)$, the angular momentum defined by the
Kepler frequency at radius $\varpi$.
In the following, we change the variable of
$\dot{M}_{\varpi\rm{,infall}}(\varpi, t)$ to relate the region where gas
accretes from the cloud core to the initial position of the gas.
We change the variable of $\dot{M}_{\varpi\rm{,infall}}(\varpi, t)$ 
from $\varpi$ to $j(\varpi)$:
\begin{equation}
 \frac{\partial}{\partial\varpi}\dot{M}_{\varpi\rm{,infall}}(\varpi,t)=\frac{\partial j}{\partial \varpi} \frac{\partial}{\partial j}\dot{M}_{\varpi\rm{,infall}}(j,t).
\label{eq:accretion_rate1}
\end{equation}
The specific angular momentum of the accreting gas is given by the
initial radius of the shell $r_{\rm{ini}}(t)$ and the angle between the initial
position of the gas and the rotational axis $\theta$,
$j=(r_{\rm{ini}}(t)\sin{\theta})^2\Omega_0$. 
Therefore, we change the variable of $\dot{M}_{\varpi\rm{,infall}}(j,t)$ again
from $j$ to $\theta$,
\begin{equation}
 \frac{\partial}{\partial j}\dot{M}_{\varpi\rm{,infall}}(j,t)=2\frac{\partial \theta}{\partial j} \frac{\partial}{\partial \theta}\dot{M}_{\varpi\rm{,infall}}(\theta,t),
\label{eq:accretion_rate2}
\end{equation}
where the factor two
comes from the fact that gas accretes on both sides of the disk from
angle $\theta$ and $\pi-\theta$ (Figure \ref{fig:cloud-disk2}).

Let us define the total mass accretion rate onto the disk, 
$\dot{M}_{\rm{infall}}(t)$, with the integral of the mass accretion rate per unit angle $\theta$,
\begin{eqnarray}
 \dot{M}_{\rm{infall}}&=& 2\int^{\frac{\pi}{2}}_{0}
  \frac{\partial\dot{M}_{\varpi\rm{,infall}}}{\partial\theta} d\theta \nonumber \\
                      &=& 2\int^{\frac{\pi}{2}}_0 2\pi\rho
		       r^2_{\rm{ini}} \frac{d r_{\rm{ini}}}{dt}
		       \sin{\theta}d\theta \nonumber \\
                      &=&4\pi \rho r^2_{\rm{ini}} \frac{dr_{\rm{ini}}}{dt},\label{Mdot_infall}
\end{eqnarray} 
where the second line is given by the total mass contained in the spherical
shell whose radius is $r_{\rm ini}$ and thickness is $dr_{\rm ini}/dt$.
Thus we can write 
\begin{equation}
 \frac{\partial \dot{M}_{\rm{\varpi,infall}}}{\partial\theta}=2\pi\rho r^2_{\rm{ini}}
\sin\theta
\frac{dr_{\rm{ini}}}{dt}=\frac{\sin\theta\dot{M}_{\rm{infall}}}{2}.
\label{eq:delMdot_deltheta}
\end{equation}
From Equations (\ref{eq:accretion_rate1}), (\ref{eq:accretion_rate2}) and (\ref{eq:delMdot_deltheta}),
we obtain a mass accretion rate of 
\begin{eqnarray}
\frac{\partial \dot{M}_{\varpi\rm{,infall}}}{\partial
 \varpi}&=&2\frac{\partial\dot{M}_{\varpi\rm{,infall}}}{\partial
 \theta}\frac{\partial
 \theta}{\partial j}\frac{\partial j}{\partial \varpi} \nonumber \\
&=&\frac{\dot{M}_{\rm{infall}}}{2\Omega_0 r^2_{\rm{ini}}}\left(1-\frac{j}{\Omega_0 r^2_{\rm{ini}}}\right)^{-\frac{1}{2}}\frac{\partial j}{\partial \varpi}.\label{eq:mass_accretion}
\end{eqnarray}

\subsection{The Evolution of Protoplanetary Disks}
Using the mass accretion rate given by
Equation (\ref{eq:mass_accretion}),
we derive equations for the evolution of protoplanetary disks \cite[cf.][]{1981Icar...48..353C}.
The angular momentum is mainly redistributed by the action of
gravitational torques in the massive disk during its early formation. 
In this work, we use a simplified one-dimensional accretion disk model
that takes into account the infall of gas from the envelope onto the disk and the transfer of angular momentum within the disk in terms of effective viscosity.
The surface density evolves according to the mass and angular momentum
conservation equations: 
\begin{equation}
\frac{\partial }{\partial t}(2\pi \varpi \Sigma)+ \frac{\partial F}{\partial
 \varpi}=\frac{d\dot{M}_{\rm{infall}}}{d\varpi},
\label{eq:disk_mass_cons}
\end{equation}
\begin{eqnarray}
\frac{\partial}{\partial t}(2\pi \varpi \Sigma j) + \frac{\partial}{\partial \varpi}(Fj) 
&= &\frac{\partial}{\partial \varpi}\left( (2\pi \varpi \Sigma) \nu
				    \varpi^2
				    \frac{\partial\Omega}{\partial
				    \varpi} \right) \nonumber \\
&&+\frac{d\dot{M}_{\rm{infall}}}{d\varpi}j_{\rm{acc}},
\label{eq:disk_angular_momentum_cons}
\end{eqnarray}
where $\Sigma$ is the surface density, $F=2\pi \varpi \Sigma v_\varpi$
is the mass flux in
the disk (where $v_\varpi$ is the radial velocity of the gas in the disk), $j$ and
$\Omega$ are the specific angular momentum and angular frequency of the
disk, and $j_{\rm{acc}}$ is the specific angular momentum accreting from the
cloud core.
We ignore the small radial pressure and radial velocity and assume
instantaneous centrifugal balance, 
\begin{equation}
\frac{j^2}{\varpi^3}=\frac{\partial \Phi}{\partial \varpi},
\label{eq:PGbalance}
\end{equation}
where $\Phi$ is the gravitational potential, approximately given by 
\begin{equation}
\frac{\partial \Phi}{\partial \varpi} = \frac{GM_\varpi}{\varpi^2},
\label{eq:gravity}
\end{equation}
where $M_\varpi$ is the sum of the mass of the central star and the disk mass
within $\varpi$.\footnote{
We measured the error of this approximation with respect to the
numerical calculation of the gravity in thin disk approximation, and
found that the difference  between our approximate calculation and thin disk
approximation was less than $14\%$ at the final snapshot when all of gas accreted on to the disk. 
}
The specific angular momentum of the gas accreting from the cloud
core is equal to that of the gas in the disk, $j_{\rm{acc}}=j$.

\begin{figure}
\epsscale{1}\plotone{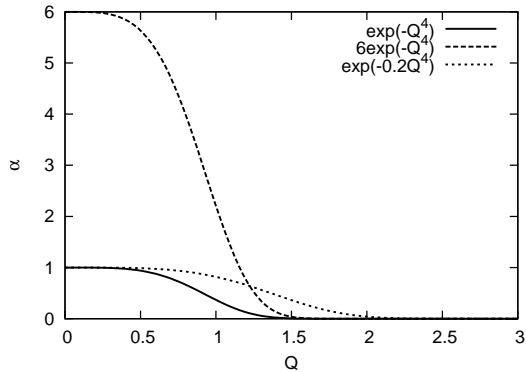}
  \caption{Gravitational
 torque parameter $\alpha=A\exp(-BQ^4)$ for $(A,B)=(1,1)$ (solid
 line), $(A,B)=(6,1)$ (dashed line), and $(A,B)=(1,0.2)$ (dotted line).
    }
  \label{fig:alpha_gra}
 \end{figure}

For the disk viscosity $\nu$, we use
\begin{equation}
\nu = \alpha c_{\rm{s}}^2/\Omega\label{eq:alpha}
\end{equation}
\cite[][]{1973A&A....24..337S}.
Since gravitational torque is effective only in gravitationally
unstable disks, $\alpha$ is large when the Toomre
parameter is $Q\lesssim2$.
In this work, and as done by \cite{2010ApJ...713.1143Z}, we model the
gravitational torque parameter as
\begin{equation}
\alpha=A\exp(-BQ^4),
\label{eq:alpha_gra}
\end{equation}
where $A$ and $B$ are chosen by fitting the results of three-dimensional simulations.
The coefficient $A$ is related to the efficiency of the angular momentum transfer
while $B$ is related to the maximum Q for which gravitational torque
is effective in a gravitationally unstable disk.
Figure \ref{fig:alpha_gra} shows $\alpha(Q)$ for parameters
$(A,B)=(1,1),\ (6,1),$ $(1,0.2)$.
The angular momentum in the disk is efficiently redistributed by the action of
 gravitational torques when $Q\lesssim 1.5$ for $(A,B)=(1,1)$, $(6,1)$, and
 $Q\lesssim 2$ for $(A,B)=(1,0.2)$.
Parameters $A$ and $B$ determine how gravitational torque becomes effective.
When some region in the disk is gravitationally unstable, spiral arms
are formed and propagate into gravitationally stable regions. Then the
non-axisymmetric gravitational field and pressure gradients affect the gravitationally
stable regions of the disk. Therefore, angular momentum transfer occurs
even in gravitationally stable regions if spiral arms appear in the
disk. 
We find that we have to modify the gravitational torque parameter given
 by Equation (\ref{eq:alpha_gra}) as follows:
\begin{eqnarray}
 \alpha=\left\{
\begin{array}{ll}
A\exp(-BQ^4)& {\rm for}\ \alpha_{\rm max}<0.1,\\
 A\exp(-BQ^4)+0.01& {\rm for}\  \alpha_{\rm max}\geqslant 0.1,
\end{array}
\right.\label{alpha_p001}
\end{eqnarray}
where $\alpha_{\rm max}$ is the maximum value of $\alpha$ in the disk.
Note that for the case $\alpha_{\rm max} \geqslant 0.1$ the value of
0.01 is added to better fit the   
three-dimensional simulations. 
The validity of this modeling is described in Section \ref{sec:3D}.

We estimate the sound speed, $c_{\rm{s}}$, to quantify the viscosity of the $\alpha$ disk model. 
To mimic the temperature evolution,  
we adopt the piecewise
polytropic equation of state
$P=\kappa\rho^\gamma$, where
\begin{eqnarray}
\gamma=\left\{
\begin{array}{ll}
1& {\rm for}\ \rho<\rho_{\rm{cri}}=2\times10^{-14}\mathrm{g\ cm^{-3}},\\
\frac{7}{5}&{\rm for}\ \rho \geqslant \rho_{\rm{cri}},
\end{array}
\right.\label{piecewise_polytrope}
\end{eqnarray}
where $\kappa=c_{{\rm{s}},10\rm{K}}^2\rho_{\rm{cri}}^{1-\gamma}$ and
$c_{{\rm{s}},10\rm{K}}=1.9\times10^4{\rm cm\  s^{-1}}$
(\cite{2006MNRAS.367...32W}; \cite{2009MNRAS.400.1563S};
\cite{2010A&A...510L...3C}; \cite{2013ApJ...763....6T}).
To evaluate the sound speed
$c_{\rm{s}}=(\gamma\kappa\rho^{\gamma-1})^{\frac{1}{2}}$ in the disk, we
need typical density at a radius $\varpi$.
When we assume a thin disk, $z\ll r$, the equation of hydrostatic
equilibrium in the $z$-direction is given by
\begin{equation}
\frac{1}{\rho}\frac{\partial P}{\partial z} = -\frac{GM}{\varpi^2}\frac{z}{\varpi}.
\end{equation}
From this equation, the density profile in the $z$-direction is given by
\begin{equation}
 \rho = \rho_0 \exp\left(-\frac{z^2}{2H^2}\right) ,
\end{equation}
where $\rho_0$ is the density at the equatorial plane and $H$ is the
scale height, $H=c_{\rm{s}}/\Omega$. By integrating 
this equation in the $z$-direction, we obtain the disk surface density
\begin{equation}
 \Sigma = \sqrt{2\pi}\rho_0 H.
\end{equation}
Thus the equation for the sound speed at the equatorial plane is 
\begin{eqnarray}
 c_{\rm{s}}^2 = \frac{dP}{d\rho}(z=0)&=& \gamma \kappa
  \rho_0^{\gamma-1}\nonumber \\
&=&\gamma\kappa \left(\frac{\Sigma}{\sqrt{2\pi}H}\right)^{\gamma-1}
 \nonumber \\
&=&\gamma\kappa \left(\frac{\Sigma\Omega}{\sqrt{2\pi}c_{\rm{s}}}\right)^{\gamma-1}.
\end{eqnarray}
Using the surface density, $\Sigma$, and angular frequency, $\Omega$,
derived in Equations (\ref{eq:disk_mass_cons}) and
(\ref{eq:disk_angular_momentum_cons}), we can estimate the sound speed
as 
\begin{equation}
c_{\rm{s}}=\left[\kappa\gamma\left(\frac{\Sigma \Omega}{\sqrt{2\pi}}\right)^{\gamma-1}\right]^{1/(\gamma+1)}.
\label{eq:c_s}
\end{equation}
\subsection{Numerical Procedures}
We calculate the formation and evolution of protoplanetary
 disks using a one-dimensional accretion disk model.
 We solve Equations (\ref{eq:disk_mass_cons}) and
 (\ref{eq:disk_angular_momentum_cons}) numerically.
We start our numerical integration with a protostar mass of 
$M_{\rm{p}}=10^{-2}M_{\odot}$ without a disk.
This initial protostellar mass corresponds to the mass of a first
core \cite[]{2000ApJ...531..350M}.
The time $t$=0 is the instant at which a core begins to collapse.
 We assume a disk radius from 0.1AU to 10000AU in the computational domain.
 The disk radius is divided into 110 logarithmically equal intervals. (We divide the
computational domain into 10 equal intervals from 0.1AU to 1AU and 100 logarithmically equal
intervals from 1AU to 10000AU.)
We assume a zero-torque condition at the center of the disk and zero-flux at
the outer boundary.
We assume that the gas inside the radius 0.1AU accretes onto the protostar.
We neglect the region of the disk from 0.1AU to 1AU because this region weakly
depends on the inner boundary condition. 
 We confirmed that an extension of the size of the computational domain
or an increase of the size and number
of grids do not significantly affect the disk evolution in the region
from 1AU to 10000AU.
\section{Results}
\label{sec:res}
\subsection{Time Evolution and Dependence on Modeling of
  Effective Viscosity}
\begin{figure}
\epsscale{1}\plotone{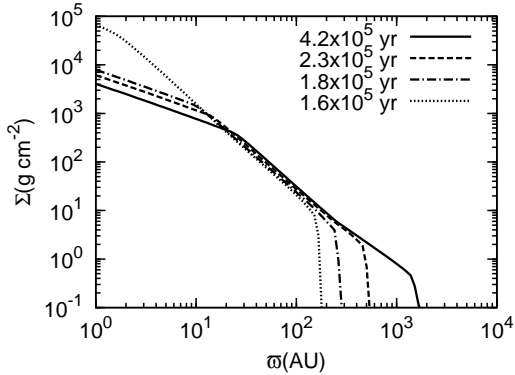}
  \caption{Surface density against resultant disk radius
$1.6\times 10^5,\ 1.8\times10^5,\ 2.3\times 10^5,\ {\rm and }\ 4.2\times10^5$
 yr after  the cloud core begins to collapse.
      }
  \label{fig:sigma_evo}
 \end{figure}

According to the prescription described in Section \ref{sec:eqs},
we calculated the evolution of surface density structures of protoplanetary disks.
Figure \ref{fig:sigma_evo} shows the time evolution of the surface density of
the resultant disk with parameters of $(A,B)=(1,1)$.
In the early stage of gas infall from the cloud core,
infalling gas has a small specific angular momentum since the initial
radius of the infalling shell is small.
As gas falls onto the disk from outer radii, the disk radius gradually increases.
The surface density decreases in the inner region because of
 viscous diffusion.
This tendency is the same as in self-similar solutions of disk evolution \cite[]{1974MNRAS.168..603L}.

\begin{figure}
\epsscale{1}\plotone{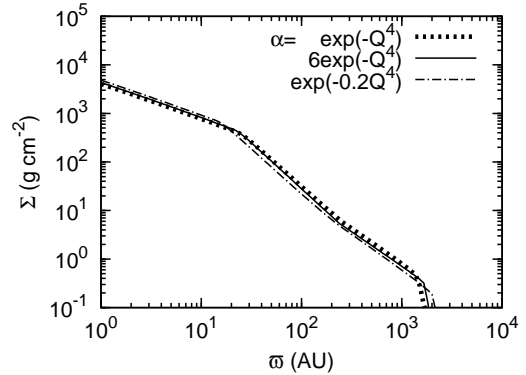}
  \caption{Surface density after all the gas of the cloud core has accreted
 onto the disk ($t=4.2\times 10^5$yr) for parameters
 $(A,B)=(1,1)$, $(A,B)=(6,1)$, and $(A,B)=(1,0.2)$.
      }
  \label{fig:sigma_prof}
 \end{figure}

Figure \ref{fig:sigma_prof} shows the surface density profile after all the gas
of the cloud core has accreted onto the disk ($t=4.2\times 10^5$yr) for three
different $\alpha$ models.
There is no significant difference in the surface densities of these models.
This suggests that the disk evolution does not depend sensitively on the
details of
modeling the effective viscosity if the angular momentum in the disk
is redistributed by gravitational torques due to gravitational instability.

 \subsection{Convergence to Self-similar Solution}
In this section we discuss the surface density profile of the resultant disk.
The disk can be divided into three regions (inner, intermediate, and
outer). 
The surface density distribution in each region converges to the self-similar solution (or static solution),
which does not account for the effects of accretion from the cloud core, as
explained below.
\subsubsection{Inner Region  $(\varpi\lesssim 20\rm{AU})$}
\begin{figure}
\epsscale{1}\plotone{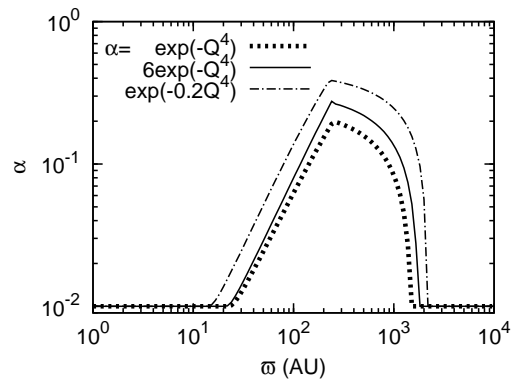}
  \caption{
Distribution of the viscous parameter $\alpha$ after all the gas of the
 cloud core has accreted onto the disk ($t=4.2\times 10^5$yr) for 
 parameters $(A,B)=(1,1)$, $(A,B)=(6,1)$, and $(A,B)=(1,0.2)$.
The parameter $\alpha$ is approximately constant ($\alpha \simeq 0.01$)
 in the region $\varpi\lesssim$20AU.
}
  \label{fig:alpha_prof}
 \end{figure}

In the inner region ($\varpi\lesssim 20$AU), the disk tends to remain gravitationally
stable with no magnetic field
(with magnetic field, this region may be gravitationally unstable;
see \citealt{2011ApJ...729...42M}). 
In this region, angular momentum is
redistributed by the action of non-axisymmetric gravity that is
caused by spiral arms formed in other regions. 
In the inner region, angular frequency and surface density tend to have
large values because this region is close to the central star. From
Equation (\ref{eq:c_s}), the sound speed is also large,
because the density is larger than the critical density (Equation (\ref{piecewise_polytrope})).
As a result, $Q$  is much larger than unity and we have $\alpha = 0.01$.
Figure \ref{fig:alpha_prof} shows the distribution of the viscous
parameter $\alpha $ at
$t=4.2\times10^5$ yr after all the
gas of the cloud core has accreted onto the disk.
The parameter $\alpha$ is approximately constant ($\alpha \simeq 0.01$) in $\varpi\lesssim$20AU.

\begin{figure}
\epsscale{1}\plotone{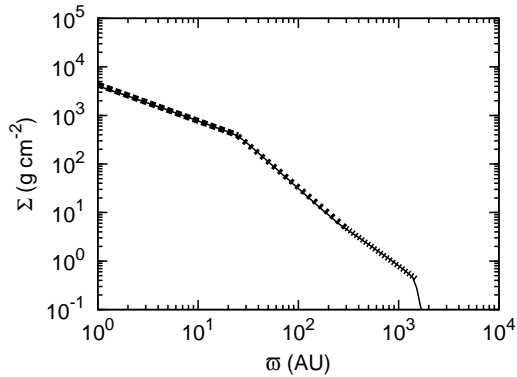}
  \caption{Surface density (solid line) derived with parameters $(A,B)=(1,1)$.
The self-similar solution (dotted line) is also plotted.
Our numerical result is in good agreement with the self-similar solution.
The surface density can be fitted by
$\Sigma \propto \varpi^{-3/4}\  (\varpi\lesssim 20 {\rm \ AU})$, 
$\Sigma \propto \varpi^{-1.76}\  (20 {\rm \ AU} \lesssim \varpi \lesssim
 300 {\rm \ AU})$, and $\Sigma \propto \varpi ^{-3/2}\  (\varpi
 \gtrsim{\rm 300\ AU}) $, respectively.
      }
  \label{fig:sigma_prof_ana}
 \end{figure}

Since we adopt a polytropic index of $\gamma=7/5$ in the inner region,
the viscosity, $\nu$, is described as
\begin{eqnarray}
\nu &=& \alpha c_{\rm s}^2/\Omega \nonumber\\
    &=&
     \frac{0.01}{\Omega}\left[\frac{7}{5}\kappa\left(\frac{\Sigma\Omega}{\sqrt{2\pi}}\right)^{\frac{2}{5}}\right]^{\frac{5}{6}}  \nonumber\\
    &\propto& \Omega^{-2/3}\Sigma^{1/3}.
\label{eq:nu_prop}
\end{eqnarray}
Suppose that angular frequency is given by Keplerian rotation 
$\Omega = \sqrt{GM_\varpi/\varpi^3}\propto \varpi^{-3/2}$. Then
Equation (\ref{eq:nu_prop}) is rewritten as
\begin{equation}
\nu\propto \varpi\Sigma^{1/3}. 
\label{eq:nu_prop2}
\end{equation}
When the surface density shows a convergence to the self-similar
solution \cite[]{1974MNRAS.168..603L},
the surface density distribution is given by
\begin{equation}
 \Sigma \propto \frac{\Omega}{\nu\varpi\frac{d\Omega}{d\varpi}}\propto
  \frac{1}{\nu}. \label{eq:sss_sigma_nu}
\end{equation}
From Equations (\ref{eq:nu_prop2}) and (\ref{eq:sss_sigma_nu}), we obtain the surface density distribution
analytically:
\begin{equation}
 \Sigma\propto \varpi^{-3/4}.
\end{equation}
Figure \ref{fig:sigma_prof_ana} shows the surface density distribution
with parameters $(A,B)=(1,1)$ and the analytic solution of $\Sigma \propto
\varpi^{-3/4}$ that fits the numerical result.
The figure indicates good agreement between the numerical calculation and
self-similar solution in the region $\varpi\lesssim 20$AU.
\subsubsection{Intermediate Region $(20\rm{AU} \lesssim \varpi \lesssim 300\rm{AU})$}
\begin{figure}
\epsscale{1}\plotone{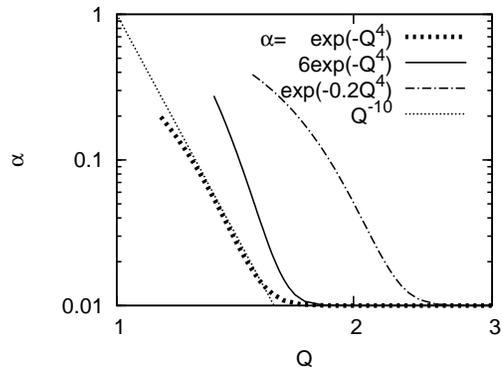}
  \caption{
Relation between the parameters
$\alpha$ and $Q$ in the disk after all of the
gas has accreted onto the disk.
The parameter $\alpha$ is roughly proportional to $Q^{-10}$ where the disk is
 gravitationally unstable with $\alpha > 0.01$.
}
  \label{fig:Q-alpha2}
 \end{figure}

In the intermediate region ($20\rm{AU} \lesssim \varpi \lesssim
300\rm{AU}$)
the disk  is gravitationally unstable ($Q \lesssim 2$) owing to a
smaller angular frequency, $\Omega$, and sound speed, $c_{\rm s}$. Thus $\alpha$ is
larger than 0.01 (Equation (\ref{alpha_p001})). 
The gas still behaves adiabatically in this region.
Figure \ref{fig:Q-alpha2} shows the relation between the parameters
$\alpha$ and $Q$ for the resultant disk after all of the
gas has accreted onto the disk.
In this region, the parameter $\alpha$ is related to $Q$ by $\alpha \propto Q^{-10}$ 
for any parameters $(A,B)$.
Since the disk mass within the radius $\varpi$ in the intermediate
region is comparable to
the central star mass, we approximate 
$\Omega =\sqrt{GM_{\varpi}/\varpi^3}\sim\sqrt{G\Sigma/\varpi} $.
Therefore the viscosity is described by
\begin{eqnarray}
 \nu&\propto& Q^{-10} c_{\rm s}^2 / \Omega \nonumber \\
     &\propto& \Sigma^{5/2}\varpi^{36/7}.
\end{eqnarray}
From Equation (\ref{eq:sss_sigma_nu}), we get $\Sigma\propto \varpi^{-1.76}$
which is in good agreement with the result of the numerical  
calculation (see Figure \ref{fig:sigma_prof_ana}).
To derive the surface density profile the relation $\alpha \propto
 Q^{-10}$ is adopted. 
 The dependence of the surface density profile on the relation between
 $\alpha$ and $Q$ is discussed in
Section \ref{sec:disc}.

\subsubsection{Outer Region $(\varpi\gtrsim300\rm{AU})$}
In the outer region ($\varpi\gtrsim300$AU), the gas behaves isothermally
because of the low surface density.
In this region the disk is still gravitationally unstable (see Figure
\ref{fig:alpha_prof}) and the Toomre parameter $Q$ is proportional to
$\Omega /\Sigma$
since the sound speed, $c_{\rm s}$, is constant.
From Equation 
(\ref{eq:sss_sigma_nu}) we obtain
\begin{equation}
 \Sigma \propto \frac{\Omega}{\alpha},
\end{equation}
where $\alpha$ is a function of $\Omega / \Sigma$.
This equation indicates that the surface density and the angular frequency have
the same radial dependence.
In this region, surface density is low such that
$M_\varpi \sim$ const.
Therefore 
$\Sigma\propto\Omega = \sqrt{GM_\varpi/\varpi^3}\sim \sqrt{GM_{tot}/\varpi^3}\propto\varpi^{-3/2}$.
Since the Toomre parameter $Q\propto \Omega/ \Sigma = \rm{const}$, the
radial dependence of the surface density distribution is 
estimated directly from the definition
of $Q$,
\begin{eqnarray}
\Sigma&=&\frac{\Omega c_{\rm s}}{\pi G Q}\\
          & \sim &1{\rm [g \ cm^{-2}]}\left(\frac{M_{tot}}{1M_\odot}\right)^{1/2}\left(\frac{\varpi}{1000\rm{AU}}\right)^{-3/2}.
\end{eqnarray}
This surface density distribution shows good agreement with the
numerical result (Figure \ref{fig:sigma_prof_ana}).

From our results, the overall surface density distribution of the
resultant disk is given by
\begin{equation}
 \Sigma \propto \left\{
\begin{array}{ll}
 \varpi ^{-3/4}& (\varpi \lesssim 20{\rm AU})\\
 \varpi ^{-1.76}&  (20{\rm AU} \lesssim \varpi \lesssim 300{\rm AU})\\
 \varpi ^{-3/2}& (\varpi \gtrsim 300 {\rm AU}).
\end{array}
\right.
\end{equation}
\section{Comparison with 3D simulations
}
\label{sec:3D}

\subsection{Model
}
To verify the evolution of the
protoplanetary disk calculated in Section \ref{sec:res}, we compare the resultant disk of our model
with that of three-dimensional hydrodynamical simulations,
originally performed in \cite{2010ApJ...724.1006M}.
\subsubsection{Basic Equations}
We solve equation of the mass conservation and 
the equation of motion including self-gravity:
\begin{eqnarray}
\frac{\partial \rho}{\partial t} + \nabla \cdot (\rho \bf{ v})&=&0,\\
\rho\frac{\partial \bf{v}}{\partial t} + \rho(\bf{v} \cdot \nabla ) \bf{v} &=& -\nabla P -\rho\nabla \phi.
\end{eqnarray}
The gravitational potential is composed of two parts:
\begin{eqnarray}
\phi &= &\phi_{\rm gas} + \phi_{\rm ps},\\
\nabla^2 \phi_{\rm gas}&=&4\pi G \rho,\\
\phi_{\rm ps}&=&-\frac{GM_{\rm ps}}{r},
\end{eqnarray}
where the subscript ps refers to protostar quantities.
We assume a barotropic gas that mimics the thermal evolution of
radiation hydrodynamical simulation \cite[]{2000ApJ...531..350M,2013ApJ...763....6T},
\begin{eqnarray}
P&=&c_{\rm s}^2\rho+\kappa\rho^\gamma \left[
				       \tanh\left(\frac{\rho}{\rho_{\rm cri}} \right) \right] ^{\frac{1}{10}},\\
\kappa &=& c_{\rm s}^2\rho_{\rm cri}^{1-\gamma},
\end{eqnarray}
where $\gamma=7/5$ and $\rho_{\rm cri}=2\times10^{-14}\mathrm{g\ cm^{-2}}$
are adopted.
When $\rho <\rho_{\rm cri}$, the gas is isothermal and the sound speed is
$c_{\rm s}=1.9\times10^4{\rm cm\ s^{-1}}$ .
When $\rho > \rho_{\rm cri}$, the gas is adiabatic with $\gamma=7/5$.
The initial condition is the same as that given in Section 2, i.e., a Bonner-Ebert
sphere whose central density is $3\times 10^5 \mathrm{cm^{-3}}$ and
radius is 17400AU.   
\subsubsection{Sink Cell}
To realize a long-term calculation of the protoplanetary disk, we adopt
a sink cell at the center of the disk. 
When the number density in the region $r < r_{\rm{sink}}=$1AU exceeds $n_{\rm{th}}=10^{12} \mathrm{cm^{-3}}$, 
we assume that the gas accretes onto the protostar, and
remove the gas exceeding the threshold density $n_{\rm th}$ from the
computational domain and add it to the protostellar mass.
We confirmed that the calculation results do not significantly depend on
the values of
$r_{\rm{sink}}$ and $n_{\rm{th}}$.

\subsection{Comparison
}
\begin{figure}
\epsscale{1}\plotone{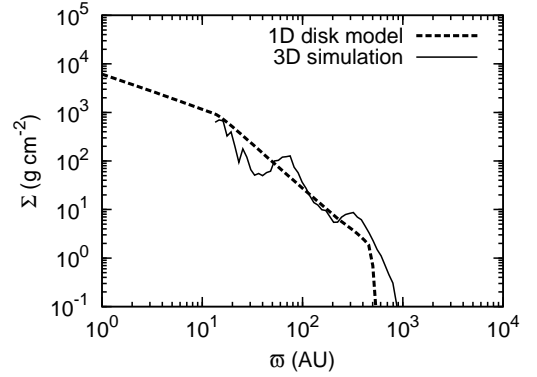}
  \caption{Surface density profiles of
resultant disks from three-dimensional simulation (solid) and one-dimensional
accretion disk model (dashed) at $t=2.3\times 10^5$ yr. 
We use $\alpha = \exp(-Q^4)$ for an effective viscosity model.
}
  \label{fig:sigma_hikaku}
 \end{figure}

Figure \ref{fig:sigma_hikaku} shows surface density profiles of the 
resultant disks from the three-dimensional simulation and one-dimensional
accretion disk model. 
We use $\alpha = \exp(-Q^4)$ for an effective viscosity model.
The surface density distribution for the effective
viscosity model is very similar to that of the three-dimensional simulation.

\begin{figure}
\epsscale{1}\plotone{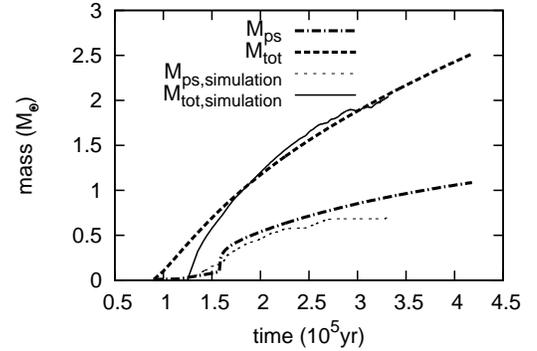}
  \caption{ Protostar and protoplanetary disk mass as a function of time
 after the
 cloud begins to collapse.
  }
  \label{fig:M-te-Q4p001_with_sim}
 \end{figure}

Figure \ref{fig:M-te-Q4p001_with_sim} shows the evolution of the 
 protostar and protoplanetary disk masses.
The mass evolution of the protostar and protoplanetary disk derived in our
effective viscosity model is similar to that of the three-dimensional
simulations.
These results show that our effective viscosity model
may provide a simplified description for the evolution of a gravitationally unstable disk formed
through the collapse of molecular a cloud core.
\section{Discussion}
\label{sec:disc}

\subsection{Dependence on Effective Viscosity
}
In this work angular momentum transfer due to gravitational torques
is modeled by $\alpha = A \exp( -BQ^{4})$ in the intermediate and outer
disk regions.
In Section \ref{sec:res} we found that the resultant disks do not depend on the
parameters $A$ and $B$ sensitively. 
We performed numerical calculations for a wide range of $A$ in order to investigate the dependence on the
model parameters.

\begin{figure}
\epsscale{1}\plotone{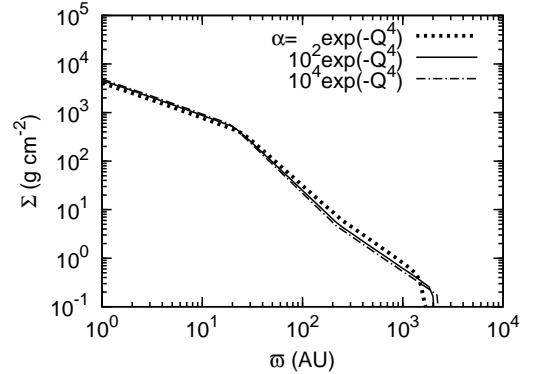}
  \caption{ 
Surface density distributions for parameters $A=1$ (dotted),
 $A=10^2$ (solid), and $A=10^4$ (dash-dotted) as a function of distance
 from the central star, $\varpi$.
  }
  \label{fig:comp_sigmaA}
 \end{figure}

\begin{figure}
\epsscale{1}\plotone{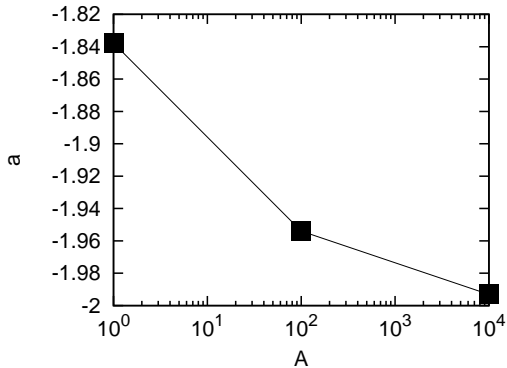}
  \caption{Relation between the parameter $A$ and the slope of the surface
 density in the intermediate region. The parameter $a$ is the exponent of surface density $\Sigma \propto \varpi^{a}$.
  }
  \label{fig:comp_slope_A}
 \end{figure}

Figure \ref{fig:comp_sigmaA} shows the surface density distributions of the resultant disks for
$A=1,\  10^2, \ {\rm and}\  10^4$. 
Although $A$ differs by four orders of magnitude, 
there is no significant difference among the models.
One of the interesting features is the slope of the surface density profile
in the intermediate region.
Figure \ref{fig:comp_slope_A} shows the relation between parameter $A$
and the slope of the surface density.
Although the slope is steeper for larger $A$, its dependence is
not so strong.
In the case
$\Sigma \propto \varpi^a$ and $\alpha \propto Q^{-b}$,
$a$ is related to $b$ by
\begin{equation}
a=-\frac{7b+4}{3b+12},
\end{equation}
in the case that the disk converges to the self-similar solutions.
Thus, if $b$ is large and $\alpha$ depends sensitively on $Q$,
the density profiles converge to $\Sigma \propto \varpi
^{-\frac{7}{3}}$.
Here, large $A$ corresponds to large $b$. 
Therefore, the value of the parameter $a$ is close to $-7/3$ when
the parameter $A$ is large as shown in Figure \ref{fig:comp_slope_A}.

\subsection{Relation between $\alpha$, $Q$, and $F$
}
\begin{figure*}
\epsscale{1.0}\plotone{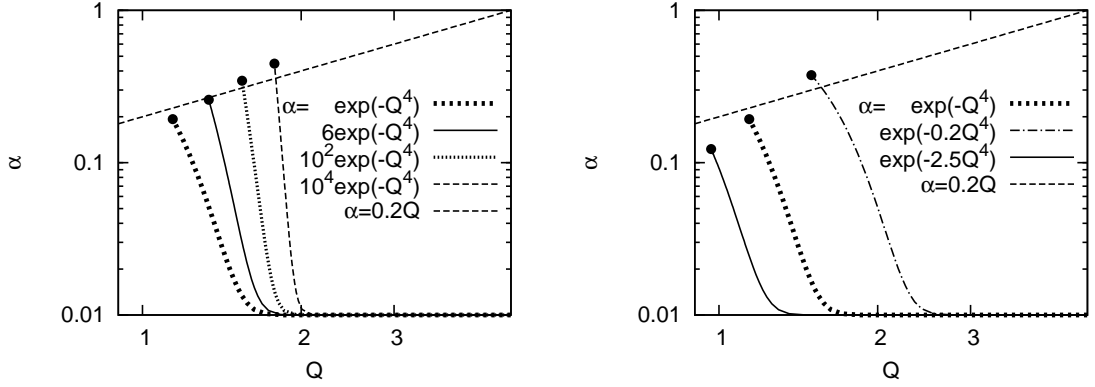}
  \caption{
  Relation between $\alpha$ and $Q$. The dotted line shows $\alpha = 0.2Q$
as given by Equation (\ref{eq:alpha_max}). Filled circles show $\alpha$
 and $Q$ at $\varpi=300$ AU.
Equation (\ref{eq:alpha_max}) can roughly estimate the value of $\alpha$
 at 300AU.
  }
  \label{fig:Q-alpha3}
 \end{figure*}

The maximum value of $\alpha$ due to gravitational torque is still
unknown.
In our model, the parameter $A$ is related to the efficiency of the angular
momentum transfer through gravitational torques.
However, Figures \ref{fig:alpha_prof} and \ref{fig:Q-alpha2}
suggest that the maximum value of $\alpha$ does not
strongly depend on $A$. The maximum value of the parameter $\alpha$
is about 0.2 for all models.
This maximum value is not controlled by the parameter $A$, but by the
mass accretion rate from the cloud core onto the disk. 
The structure of the disk converges to the self-similar
solution, and the mass flux in the disk is similar to the gas
accretion rate from the cloud core to the protoplanetary disk.  
In the case that the mass flux in the disk is less than the gas 
accretion from the
cloud core, gas accumulates on the disk. Therefore, the disk becomes
gravitationally unstable; $Q$ decreases and $\alpha$ increases.
On the other hand, in the case that the mass flux is larger than the gas
accretion from the cloud core, the disk mass decreases. Then, the disk becomes
gravitationally stable and the mass flux decreases.
 As a result, the mass flux in the disk has a similar value to the mass
accretion rate from the cloud core.
Using Equation (\ref{eq:disk_mass_cons}) and (\ref{eq:disk_angular_momentum_cons}),
the mass flux in the disk is given by
\begin{equation}
 F=\frac{1}{\frac{\partial j}{\partial
  \varpi}}\left[\frac{\partial}{\partial \varpi}\left(2\pi \Sigma \nu
						 \varpi ^3
						 \frac{\partial
						 \Omega}{\partial
						 \varpi}\right)-2\pi\varpi\Sigma\frac{\partial
  j}{\partial t}\right].\label{eq:flux}
\end{equation}
We ignore the second term of Equation (\ref{eq:flux}) because this is
smaller than the first term and use the
relations 
$\Omega \propto \varpi ^{-1.5} \ {\rm and} \ \Sigma\propto\varpi^{-1.5}$ for simplicity.
Using Equation (\ref{eq:alpha}) and Toomre's $Q$
parameter, we can rewrite Equation
(\ref{eq:flux}) as
\begin{equation}
 F \simeq -3\left(\frac{c_{\rm s}^2}{v_\phi^2}\right)^{\frac{3}{2}}G^{1/2} M_\varpi^{3/2} \varpi^{-3/2}\frac{\alpha}{Q}.\label{eq:ap_flux}
\end{equation}
Using a typical mass
accretion rate from the cloud core, Equation (\ref{eq:ap_flux}) is
given by 
$F\sim c_{\rm s}^3/G\sim2\times10^{-6}M_{\odot}\ \rm{yr}^{-1}$, such that
\begin{eqnarray}
\alpha &\simeq& 0.2 Q\left(\frac{F}{2\times 10^{-6}M_{\odot}
		      \ \rm{yr}^{-1}}\right)\left(\frac{v_\phi^2/c_{\rm s}^2}{100}\right)^{\frac{3}{2}}\nonumber \\
 &&\times\left(\frac{r}{300\rm{AU}}\right)^{\frac{3}{2}}
                       \left(\frac{M_r}{1M_\odot}\right)^{-\frac{3}{2}}.                       \label{eq:alpha_max}
 \end{eqnarray}
Therefore we obtain $\alpha  = 0.2 Q$ at $\varpi=300$ AU at which
$\alpha$ has its
largest value in the disk (Figure \ref{fig:alpha_prof}).
Figure \ref{fig:Q-alpha3} shows the function $\alpha =0.2 Q$ and also
shows the relations between $\alpha $ and $Q$
after all of the gas of the cloud core accretes onto 
the disk ($t=4.2 \times 10^5$yr) for different values of the
parameters $(A,B)$.
Equation (\ref{eq:alpha_max}) can roughly estimate the value of $\alpha$
 at 300 AU.
If we know where the value of $\alpha$ is largest, Equation
  (\ref{eq:alpha_max}) gives the maximum $\alpha$. 
Thus, the maximum $\alpha$ is related to the gas accretion rate from the cloud
core, not the parameter $A$.
In other words, the parameter $\alpha$ changes only slightly even if 
the functional form of the effective $\alpha$ changes drastically.
Even if we change the parameter $A$ by four orders of magnitude,
the difference between parameters $\alpha $ and $Q$ is less than factor three
(Figure \ref{fig:Q-alpha3}).
This is because we use the effective viscosity model in which $\alpha$ changes
drastically when $Q$ changes slightly.
When model parameters for viscosity, such as $A$, are changed,
the mass flux in the disk balances the mass accretion
rate from the cloud core by changing the surface density $\Sigma$ slightly.
Therefore, the structure of the disk is self-regulated to achieve the
balance between the mass flux in the disk and the mass accretion rate from the
cloud core. Thus, the surface density of the resultant disk  does not
depend on the details of the modeling of effective viscosity. 
\subsection{Dependence on Initial Condition
}
We have shown the resultant disk formed from a cloud core appropriated
by a Bonner-Ebert sphere with mass
$2.5M_{\odot}$, angular frequency $\Omega_0=4.8\times 10^{-14}\ {\rm s^{-1}}$, and mass
enhancement factor $f=$ 1.4.
Now we change the parameters of the cloud core, and discuss the
dependence of the resultant disk on the initial conditions.
\subsubsection{Angular Momentum of Cloud Cores
}
\begin{figure}
\epsscale{1}\plotone{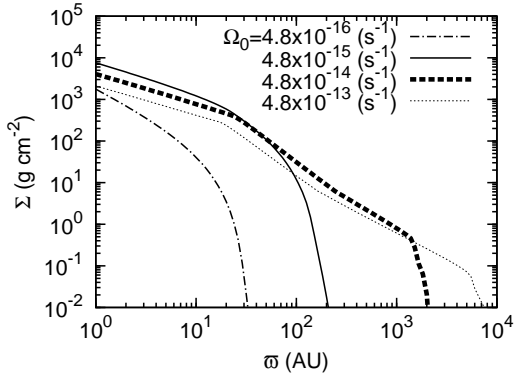}
  \caption{
Surface density distributions of the resultant disks for $\Omega
_0 = 4.8 \times 10^{-16},\ 4.8 \times 10^{-15},\ 4.8 \times 10^{-14},\
 4.8 \times 10^{-13}\  \rm s^{-1}$ after all of the gas of the cloud core
 has accreted onto the disk. 
  }
  \label{fig:sigma_dep_Omega}
 \end{figure}

It remains difficult to observe the angular
momentum of a cloud core.
However, the angular momentum of the accreting gas depends on that of the cloud core, which determines  
the radius at which the gas accretes.
Thus, the initial angular momentum of the cloud core plays an important
role in the formation of protoplanetary disks.
To investigate the effects of initial angular momentum, we performed a
number of calculations of 
protoplanetary disk formation with cloud cores of different angular frequencies.
The surface density distributions of the resultant disks are shown in
Figure \ref{fig:sigma_dep_Omega}.
Since gas accretes onto the region where the gravitational force is balanced
by the centrifugal force,
gas accretes onto the outer region of the disk when the angular
frequency of the cloud core is sufficiently large. Thus, such a core forms larger
disk.

The initial angular momentum of a cloud core does not affect the mass
accretion rate onto a disk (Equation (\ref{Mdot_infall})).
As discussed above, the order of the mass flux in the disk is the same as
the order of the mass
accretion rate from the cloud core to the disk, $\sim 10^{-6} M_{\odot}\ 
{\rm yr}^{-1}$.
The mass flux in the disk depends heavily on the surface density
through the $\alpha$ parameter.
Thus, the fact that mass flux does not depend on the angular
momentum of the cloud core means that the value of the surface density
does not depend sensitively on the
angular momentum of the cloud core. 

So far a rigidly rotating cloud core is used as the initial condition
for our fiducial models.
The rotation laws of the actual star-forming dense cores in molecular
clouds are unknown observationally. 
Thus, in the following we investigate how the result depends on the
initial rotation profile, by comparing the surface density structures of
the resultant disks formed
from cloud cores with various rotation profile, 
$\Omega_0 \propto \varpi^0,\ \varpi^{-0.2},\ \varpi^{-0.5},\
\varpi^{-1}$, where $\varpi$ is a distance from a rotation
axis. These cores have the same total angular momentum.
Since gas accretes onto the disk region where central
gravitational force is balanced by centrifugal force, angular momentum
distributions of the cloud cores are affect the region where gas accretes
from the cloud cores.  
Differentially rotating cores ($\Omega_0 \propto
\varpi^{-0.2},\ \varpi^{-0.5},\ \varpi^{-1}$) have larger angular
momenta in the inner pert and smaller angular momenta in the outer
pert than the rigidly rotating core.
Therefore disks formed from the differentially rotating cores have
larger angular
momenta than that of rigidly rotating core in the early phase of
accretion.
At the end of the accretion, gas with smaller angular momentum accretes onto the
disk from differentially rotating cores than in the rigidly rotating
core.

\begin{figure}
\epsscale{1}\plotone{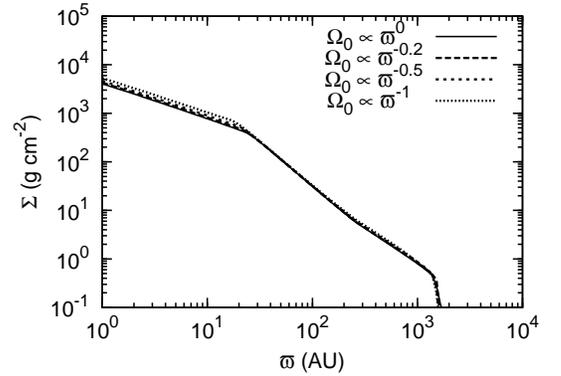}
  \caption{
Surface density distributions of the
resultant disks for various initial rotation profiles,
 $\Omega_0 \propto \varpi^0,\ \varpi^{-0.2},\
 \varpi^{-0.5},\ \varpi^{-1}$ just after all of the gas in the cloud
 core accrete onto the disk. Total angular momentum of the cloud
 cores are the same as that of our fiducial model with 
$\Omega_0 = 4.8\times 10^{-14}\ {\mathrm s^{-1}}$.
  }
  \label{fig:sigma_dep_Omega_dist}
 \end{figure}

Figure \ref{fig:sigma_dep_Omega_dist} shows the density distributions of the
resultant disks after all the gas in the cloud core accrete onto the disk.
Although the initial angular momentum distributions are different, 
the surface density
distributions of resultant disks are similar to that
from rigidly rotating core.
Since angular momentum of the disk is well redistributed by the
gravitational torque, the difference in the initial angular momentum
distribution of the cloud
core produces only a small effect on the resultant disk.

\subsubsection{Mass Accretion Rate from the Cloud Core onto the Disk
}
\begin{figure}
\epsscale{1}\plotone{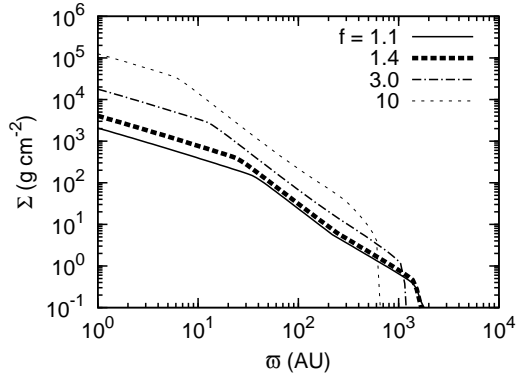}
  \caption{
Surface density distributions of the
resultant disks for $\Omega_0=4.8\times 10^{-14} {\rm s^{-1}}$ and
 $f=1.1,\  1.4,\  3,\  10$  after all of the gas of the cloud
 core has accreted onto the disk. 
  }
  \label{fig:sigma_dep_f}
 \end{figure}

\begin{figure}
\epsscale{1}\plotone{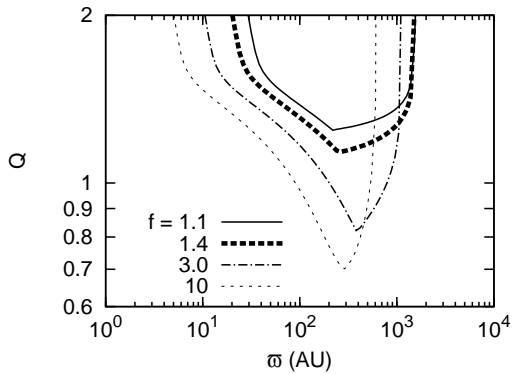}
  \caption{
Distributions of $Q$ of the resultant disks for $\Omega_0=4.8\times
 10^{-14} {\rm s^{-1}}$ and $f=1.1, 1.4, 3, 10$  after all of the gas of the cloud
 core has accreted onto the disk.
  }
  \label{fig:Q_dep_f}
 \end{figure}

The mass accretion rate from the cloud core onto the disk depends on
the properties of the cloud core. Therefore the formation
process of the cloud core should determine the properties of the cloud core.
In our fiducial model, we increased the density of the Bonnor-Ebert
spheres by a 
factor of $f=1.4$, corresponding to $U/|W|=0.5$, in order to promote 
gravitational collapse.
We can change the mass accretion rate onto the disk by changing $f$.
A large value of $f$ (this means small $U/|W|$) corresponds to
a large  mass accretion rate onto the disk.
We performed numerical calculations with $f=1.1,\  1.4,\  3,\  10$
($U/|W|=0.64,\ 0.5,\ 0.23,\ 0.07$).
Figure \ref{fig:sigma_dep_f} shows the density distributions of the
resultant disks after all the gas of the cloud core has accreted onto the disk.
When we adopt larger $f$ (or smaller $U/|W|$), both the mass accretion rate onto the disk 
and the mass flux in the disk are large.
As a result, the surface density is large for larger $f$.
On the other hand, it takes longer time for gas to accrete from a cloud
 core with smaller $f$.
 At the outer edge of the disk, the angular momentum of the gas increases
because angular momentum is transported from the inner part to the outer
part.
Thus, the disk radius increases over time.
Therefore, a larger disk forms when $f$ is small. 
Figure \ref{fig:Q_dep_f} shows the distributions of $Q$ for various $f$. 
For models with $f>3$, the Toomre parameter $Q$ is smaller than unity at the
 radius $\varpi \sim300$AU ($Q\simeq 0.8$ for $f=3$, $Q\simeq 0.7$ for
 $f=10$).

\section{Conclusions}
\label{sec:conc}
In this paper we investigated the formation process of
self-gravitating protoplanetary disks.
We developed a simplified one-dimensional accretion disk model that
accounts for the infall of gas from the envelope onto the disk. We also
modeled the transfer of angular momentum within the disk in terms of an effective viscosity. 
The resultant disk consist of three regions.
The inner region of the disk are adiabatic and gravitationally
stable. The intermediate region is also adiabatic but
gravitationally unstable. The outer region is isothermal and
gravitationally unstable. The structure of the surface density profiles of the
disks converge to self-similar solutions. The radial dependence of
the disk surface density is described by
$\Sigma \propto \varpi ^{-3/4}$ for the inner region,
$\Sigma \propto \varpi ^{-1.7}\sim\varpi^{-2}$ for the intermediate region,
 and 
$\Sigma \propto \varpi ^{-3/2}$ for the outer region.

We also performed three-dimensional numerical simulations starting from
the collapse of
cloud cores to the formation of the protoplanetary disks. 
We compared the three-dimensional simulations with our one-dimensional
accretion disk model and confirmed that our model of effective viscosity 
can provide a
simplified description for the evolution of a gravitationally unstable
disk.

In addition, we used an effective viscosity model in which $\alpha$
changes sensitively with $Q$ when the disk is gravitationally unstable.
This model shows a strong self-regulation mechanism in protoplanetary disks.

The structure of the disk depends on the initial state of the cloud
core. The disk radii depends on the initial rotation frequency of
the cloud core. Massive, gravitationally unstable cloud cores cause a high mass
accretion rate onto the disk. In such a cloud, a massive disk forms and
becomes gravitationally unstable.
In this work, $Q$ is minimum at $\varpi \sim 300$ AU which is the
boundary between the 
intermediate and outer disk regions.
This result suggests that fragmentation of the disk occurs at outer radii
$>100$AU in accordance with three-dimensional numerical simulations of 
protostellar collapse \citep[e.g.,][]{2010ApJ...724.1006M}

With our effective viscosity model, we are able to calculate protoplanetary disk
formation more easily and more rapidly than using three-dimensional simulations.
Our model is thus a useful tool for further modeling of chemistry,
radiative transfer and planet formation in protoplanetary disks.

We thank Takashi Nakamura and Kazuyuki Omukai for their 
continuous encouragement and 
Kohei Inayoshi for fruitful discussion.
Data analysis was carried out in part on the 
Yukawa Institute Computer Facility and the general-purpose PC farm
 at the Center for Computational Astrophysics, CfCA, of the National
 Astronomical Observatory of Japan.


\end{document}